# Dynamic characterization of crystalline and glass phases of deuterated 1,1,2,2 Tetrachloroethane

Silvina C. Pérez[1,a)], Mariano Zuriaga[1,b)], Pablo Serra[1,c)], Alberto Wolfenson [1,d)], Philippe Negrier[2,e)] and Josep Lluis Tamarit[3,f)]

[1]*Facultad de Matemática, Astronomía y Física, Universidad Nacional de Córdoba and IFEG-CONICET, Ciudad Universitaria, X5016LAE Córdoba, Argentina.*

[2] *Université Bordeaux, LOMA, UMR 5798, F-33400 Talence, France CNRS, LOMA, UMR 5798, F-33400 Talence, France*

[3]*Grup de Caracterització de Materials, Departament de Física i Enginyeria Nuclear, ETSEIB, Diagonal 647, Universitat Politècnica de Catalunya, 08028 Barcelona, Catalonia (Spain)*

**ABSTRACT**

A thorough characterization of the γ, β and glass phases of deuterated 1,1,2,2 Tetrachloroethane ($C_2D_2Cl_4$) via Nuclear Quadrupole Resonance and Molecular Dynamic Simulations (MDS) is reported. The presence of molecular reorientations was experimentally observed in the glass phase and in the $\beta$ phase. In the $\beta$ phase, and from MDS, these reorientations are attributed to two possible movements, i.e. a 180º reorientation around the $C_2$ molecular symmetry axis and a reorientation of the molecule between two non-equivalent positions. In the glass phase, the spin-lattice relaxation time $T_1$ is of the order of 16 times lower than in the crystalline phase and varies as $T^{-1}$ below 100 K in good agreement with the strong quadrupolar relaxation observed in amorphous materials and in the glassy state of molecular organic systems. The activation energy of molecular reorientations in the glass phase (19 kJ/mol) is comparable to that observed in the glassy crystal of a "molecular cousin" compound, Freon 112 ($C_2F_2Cl_4$), for the secondary $\beta$-relaxation. Moreover, the on-site orientational motion of Tetrachloroethane molecules offers a new indirect evidence of the prominent role of such orientational disorder in glassy dynamics.

a) clyde@famaf.unc.edu.ar
b) zuriaga@famaf.unc.edu.ar
c) serra@famaf.unc.edu.ar
d) wolf@famaf.unc.edu.ar
e) philippe.negrier@u-bordeaux.fr
f) josep.lluis.tamarit@upc.edu

**I. INTRODUCTION**

The 1,1,2,2-tetrachloroethane (TCE) molecule can exist in both trans and gauche conformations. The slight energy difference between both conformers (<1 kcal mol$^{-1}$) [1-3] means that the polymorphic phases that appear at different temperatures and pressures depend mainly on the intermolecular interactions.[4]

At high pressures (ca. 0.65 GPa) the structure of TCE (phase $\alpha$) has been found to be monoclinic (P2$_1$/c, Z = 2, Z' = 0.5) and formed by molecules with *trans* conformation only.[4] At normal pressure, the stable $\beta$ phase is known to be orthorhombic (P2$_1$2$_1$2$_1$, Z = 8, Z' = 2) and formed by molecules with one of the two gauche conformations (*gauche +*). Recently, a metastable solid form (phase $\gamma$) has been reported,[5] which is obtained by crystallization, upon heating the glass obtained after quenching of the melt.

The structure of phase $\gamma$ was found to be monoclinic (P2$_1$/c) with Z = 8 molecules in the unit cell and Z' = 2 in the asymmetric unit with the following main characteristic: in this new polymorph the two *gauche* conformers coexist in the asymmetric unit. Overlaying crystalline structures of $\beta$ and $\gamma$ phases shows that 50% of the molecules remain in essentially identical positions, while the other 50% of the molecules has their positions (approximately) related by a reflection perpendicular to the *b* axis, which means that the carbon-carbon bonds form an angle of approximately 73°. This is shown in Figure 1.

In order to dynamically characterize each of the polymorphic forms existing at normal pressures, a Nuclear Quadrupole Resonance (NQR) technique was used in this work.

It is well known that a nucleus interacts with the electronic environment, not only through magnetic hyperfine couplings due to its magnetic moment, but also through the interaction of its quadrupole moment with the local Electric Field Gradient (EFG), either statically or dynamically. The EFG arises from a non-symmetric distribution of electric charges around the nucleus. These electric charges can originate from non-bonding electrons, electrons in the bond, and charges of neighboring atoms or ions. Therefore, NQR is a highly sensitive tool for studying solids as it provides detailed information on the static and dynamic properties of the structure in the scale of a few interatomic distances. Thus, it may be regarded as a powerful tool for investigating the local order in solids,[6-13] whereas the interpretation of the data from usual scattering experiments, such as X-ray or neutron scattering, on disordered or complex-ordered materials is complicated due to the absence of long-range translational symmetry or large and complicated unit cell.[14, 15]

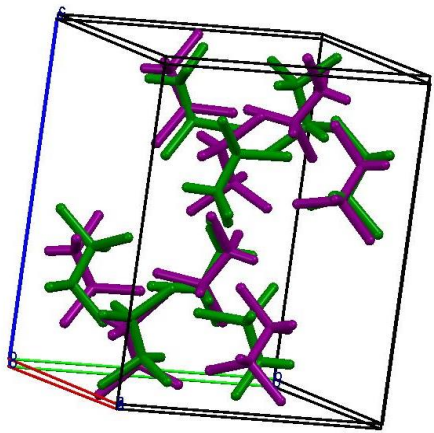

FIG. 1: Overlap of crystalline structures of $\beta$ (Green) and $\gamma$ (Purple) phases in TCE.

Moreover, NQR has proven to be a useful technique, together with the dielectric relaxation, to reveal the origin of the $\beta$-relaxation in Chlorobromomethanes.[16]

The microscopic origin of the glass transition for these materials can be considered as a benchmark to highlight the relevance of the orientational degrees of freedom due to the statistical disorder of some molecular entities. In general, canonical glasses are obtained by freezing the translational and orientational degrees of freedom present in the liquid state, which has been supercooled. The complexity of the problem can be reduced by regarding simpler systems in which only orientational degrees of freedom are frozen giving rise to orientational glasses or glassy crystals by supercooling the orientationally disordered (or plastic) phases.[17-19] The complexity can be further reduced analyzing systems in which the orientational disorder is due to the existence of an intrinsic statistical disorder involving the site occupancy of several atoms of the molecular entities. In some of these cases, it has been demonstrated that the main glass features show up, pointing out the relevance of the orientational degrees of freedom as far as the glass properties are considered. [20-22] Also, it should be stressed that for all those disordered systems the collective dynamics of the entities cannot be decoupled from the molecular or localized excitations, and thus the use of pertinent techniques, which encompass the different origins of disorder, should provide important highlights to the glass transition problem.

Therefore, in the present work, NQR studies were done in TCE, not only to characterize dynamically the three different phases of deuterated TCE, but also in an attempt to contribute to the understanding of the dynamics of Freon 112, in the same way as NQR studies of the monoclinic phase of Tetrachlorometane have contributed to the understanding of the dynamics of Chlorobromomethanes.[20] Additionally, molecular dynamics simulations (MDS) were performed in order to disclose the underlying physics and thus to obtain a clear microscopic picture of the molecular movements. Simulations have already been done on rotational motions in orientationally disordered crystals such as difluorotetrachloroethane[23, 24] and chloroadamantane.[25]

## II. EXPERIMENTAL

In NQR, the interaction of the nuclear quadrupole moment with the EFG generated by the surrounding charges is characterized by discrete energy levels which can be detected, as in NMR, by measuring the absorption (dispersion) of a signal in a resonant circuit that contains the sample studied. In this sense, this technique is equivalent to NMR.

The number of lines in the NQR spectrum is equal to the number of non-equivalent nuclei in the crystal cell and the area under the NQR line is proportional to the number of resonant nuclei at that frequency. Other important parameters are, as in NMR, the spin-lattice relaxation time ($T_1$) and the spin-spin relaxation time ($T_2$). The temperature dependence of all NQR parameters is useful in building a molecular picture of what is taking place in the crystal (phase transitions, molecular reorientations, etc...).[26]

The magnitude of the EFG at a nuclear site is an extremely sensitive function of its near environment. In a disordered crystal, for example, the configuration of nearby atoms varies from site to site and, in this way, the overall linewidth of the resonance is considerably broadened.[20, 27] In a conventional pulse spectrometer,[28] the NQR spectrum is obtained from the Fourier transform of the free induction decay (FID) which results from a radiofrequency pulse with a length of $t_w$. In disordered crystals, the line broadening shortens the length of the FID and, due to the dead time of the receiver, no signal is observed after the pulse. Therefore, the line shapes are obtained by using spin−echo Fourier transform mapping spectroscopy,[29] in which an echo Hahn sequence of two coherent radiofrequency pulses are applied with a time interval of $\tau$ between them.[30] With this method, the line shape is obtained by recording the $(\pi/2)-\tau-(\pi)$ echo signals and adding up its Fourier transforms for different irradiation frequencies. This reconstruction technique is also used when multiple resonance lines are close together in the NQR spectrum. In the present work, for phases $\beta$ and $\gamma$, the number of averages was 40, $\pi/2 = 14$ μs, and $\tau = 100$ μs; while for the glassy phase the number of averages was 1000, $\pi/2 = 14$ μs, and $\tau = 70$ μs. In all cases the mapping frequency step used was 13.6 kHz.

The $T_1$ measurements were made, upon the echo, using the standard saturation-recovery method. In this technique, a saturating $\pi/2$ pulse is initially applied. At a time $t$ after the completion of the initial pulse, an interrogation sequence $(\pi/2) - \tau - (\pi)$ is applied.

The amplitude of the half-echo Fourier transform is proportional to $(1 - e^{-t/T_1})$. In the glassy phase, $T_1$ was measured at the central frequency of the broad NQR spectrum with $\pi/2 = 18$ μs and $\tau = 50$ μs, while in phases β and γ, $T_1$ was measured for each line in the spectrum with $\pi/2 = 14$ μs and $\tau = 100$ μs.

As for $T_2$, it was obtained by observing the height of the spin echo as a function of $t$ using the Hahn echo sequence $\pi/2 - t - \pi$. The height of the echo is given by $S = S_o e^{-2t/T_2}$ [30].

The different values of $\tau$ chosen in the NQR spectrum and for $T_1$ measurements are due to the fact that $T_2$ is shorter in the glassy phase than in the ordered phases.

The compound used in the present investigation, deuterated 1,1,2,2-tetrachlorethane ($C_2Cl_4D_2$), was obtained from Aldrich Chemical Company with a purity of 99.5% and used without further purification. $^{35}Cl$ NQR frequency and spin-lattice relaxation time measurements were done using a Fourier transform pulse spectrometer with a Tecmag NMRkit II unit and a Macintosh based real-time NMR station. The sample container was a glass cylinder of 3 cm length and 1 cm diameter, closed under vacuum.

The temperature was controlled to within 0.1 K using a homemade cryogenic system with a Lakeshore temperature controller. The temperature range covered was between 80 K and the melting point of 1,1,2,2-tetrachlorethane at 229 K.

**III. MOLECULAR DYNAMICS SIMULATIONS**

MDS of TCE have been performed in β and γ phases in the NVT ensemble. Rigid molecules were considered in gauche configuration and the intramolecular parameters used were obtained from previously published X-ray diffraction data in Ref. 5. The intermolecular interactions were described by Lennard-Jones (L-J) and Coulombic potentials (see Table I).[31-34] The L-J parameters (σ, ε) for interactions between different atoms were determined using the combination rules $\sigma_{i,j} = (\sigma_i + \sigma_j)/2$ and $\varepsilon_{i,j} = \sqrt{\varepsilon_i \varepsilon_j}$, where i,j represent the atoms C, Cl, H.

**Table I:** Charge of Coulombic potential and Lennard-Jones parameters used in MDS.

|    | charge/e | σ [nm]   | ε [kJ/mol] |
|----|----------|----------|------------|
| C  | -0.2308  | 0.3397   | 0.45773    |
| H  | 0.2644   | 0.229317 | 0.656888   |
| Cl | -0.0168  | 0.347094 | 1.10876    |

The NVT MDS were performed using the Gromacs v5.0.2 package,[35] using a leap-frog algorithm with a time step of 0.0005 ps and a velocity rescale thermostat with a time constant of 2 ps. As initial configuration, the experimental volume and the perfect crystalline structure determined by X-ray diffraction were used.[4, 5]

The system was formed by 800 molecules (6400 atoms), and some tests with a larger system of 6400 molecules (51200 atoms) were done in order to discard finite-size effects. Runs of 20000 ps were done, taking averages over the last 5000 ps. Some very large runs (300000 ps) were made for some temperatures, in order to corroborate the shorter time results.

It is known from X-ray diffraction that only the gauche conformation exists in phase $\beta$.[4] Even though fast jumps gauche-trans-gauche of some molecules are not observed in X-ray diffraction, their presence could indeed modify the spin-lattice relaxation time. In order to discard this phenomenon, we performed NVT MDS with non-rigid molecules. In these simulations we included three intramolecular interactions, (a) harmonic atom-atom forces, (b) three-body harmonic angle potentials, and (c) a four-body dihedral Ryckaert-Bellemans potential allowing torsion of the molecule around the C-C bond. The parameters were obtained from references.[36-38] A MDS was performed during 150000 ps for the largest systems (6400 molecules) . No jumps or changes were observed in the molecular configuration.

**IV RESULTS AND DISCUSSION**

When liquid deuterated TCE is quenched to 77 K, a broad NQR spectrum (around 1 MHz FWHM) is observed, compatible with a glassy state. Due to the shortening of $T_2$, the NQR signal in this phase is observed for temperatures of up to ~120 K. Around 165 K, the supercooled liquid ($T_g$ is around 153 K) crystallizes and phase γ appears, characterized by eight peaks, and circa 190 K this phase transforms irreversibly to phase β, also characterized by eight NQR frequencies. The eight peaks in both phases are due to the eight Cl atoms in the irreducible unit cell (Z' = 2 for both β and γ phases). Once the γ or the β phase is obtained, it is possible to cool to the liquid nitrogen temperature (see Figure 2).

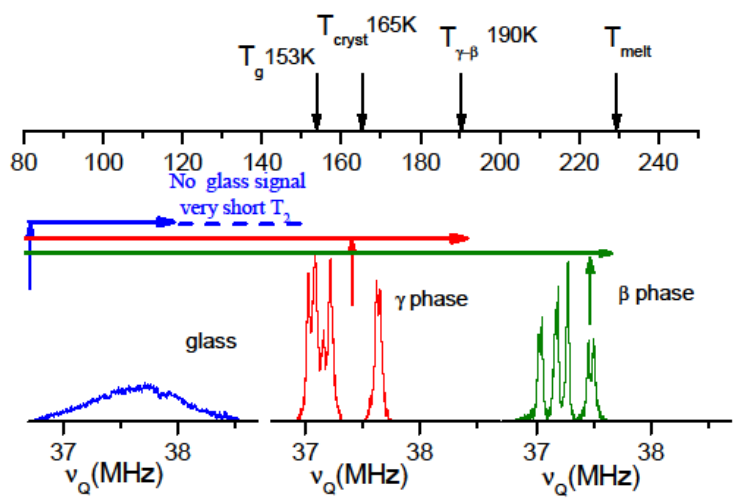

FIG. 2: NQR spectra at 80 K in the glass (left blue spectrum), β (central red spectrum) and γ (right green spectrum) phases of deuterated TCE. Transition temperatures are also indicated in the upper panel.

Figure 3 shows detailed NQR spectra for phase $\gamma$ and $\beta$ at 90 K. Each phase is characterized by eight Lorentzian peaks of similar line width, distributed in almost the same frequency range. In phase $\gamma$, the frequency range is always wider than in phase $\beta$.

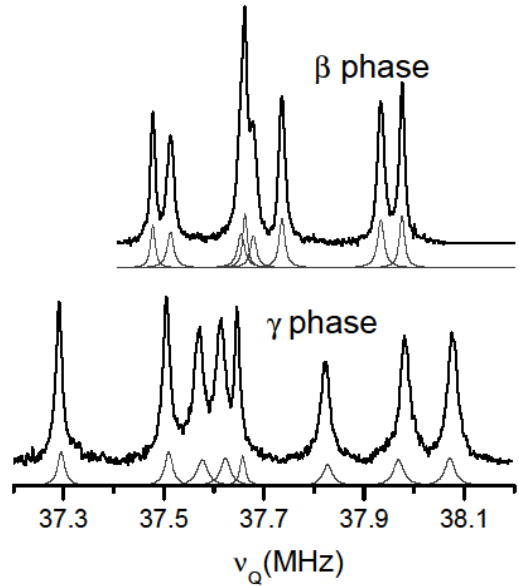

FIG. 3: NQR spectra of phases $\beta$ and $\gamma$ at 90 K The eight Lorentzian peaks used to fit the NQR spectra are shown at the bottom of each figure.

*A. NQR frequency*

Figure 4 shows the temperature dependence of the NQR frequencies in the stable phase $\beta$ and in the metastable phase $\gamma$. While the eight frequencies of phase $\beta$ decrease with temperature at approximately the same rate, the same does not occur for phase $\gamma$ where four lines (filled triangles) have a stronger temperature dependence than the other four.

Thermal motions present in all crystals occur at frequencies much higher than those of NQR transitions; thus the EFG, measured by NQR, is time-averaged over these motions. In the absence of phase transitions, the $^{35}$Cl-NQR frequency smoothly and slowly decreases with increasing temperature.

Bayer's theory assumes that the averaging effect of molecular vibrations on EFG is reflected in the NQR frequency according to:[26, 39]

$$\nu_Q = \nu_o \left(1 - \frac{3}{2}\langle\theta^2\rangle\right) \quad (1)$$

where $\nu_o$ is the NQR frequency in a rigid lattice and $\langle\theta^2\rangle$ is the mean square angular displacement of the EFG z-axis from its equilibrium position. In the majority of molecular crystals with single-bonded covalent chlorine, the z axis of the EFG lies along the C – Cl bond axis within experimental error (~1° or 2°).[40, 41] Therefore, in the present case it is possible to assume that the z-axis corresponds to the C-Cl bond.

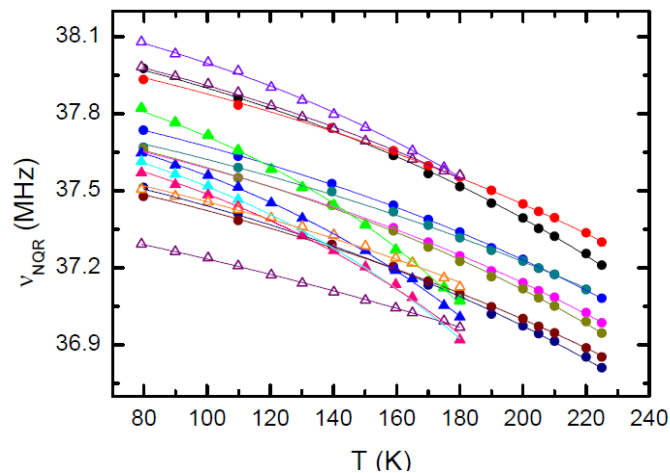

FIG. 4: $\nu_Q$ as a function of temperature in phase $\beta$ (filled circles) and in phase $\gamma$ (filled and empty triangles) of deuterated TCE.

Using Eq. (1) it is possible, from experimental data, to find the temperature dependence of $\langle\theta^2\rangle$:

$$\frac{\nu_Q(80K) - \nu_Q(T)}{\nu_Q(80K)} \approx \langle\theta^2\rangle_T - \langle\theta^2\rangle_{80K} = \Delta\theta^2 \quad (2)$$

Figure 5a shows this behavior for the eight lines in each phase. It is observed that in phase $\gamma$ there are two $\Delta\theta^2$ groups, one of them being similar to $\Delta\theta^2$ in phase $\beta$. Two explanations are possible: a) librations of the two non-equivalent molecules are different because of the differences between $\beta$ and $\gamma$ phases shown in Figure 1 or b) $\langle \theta^2 \rangle$ for different Cl atoms in the same molecule are not equal.

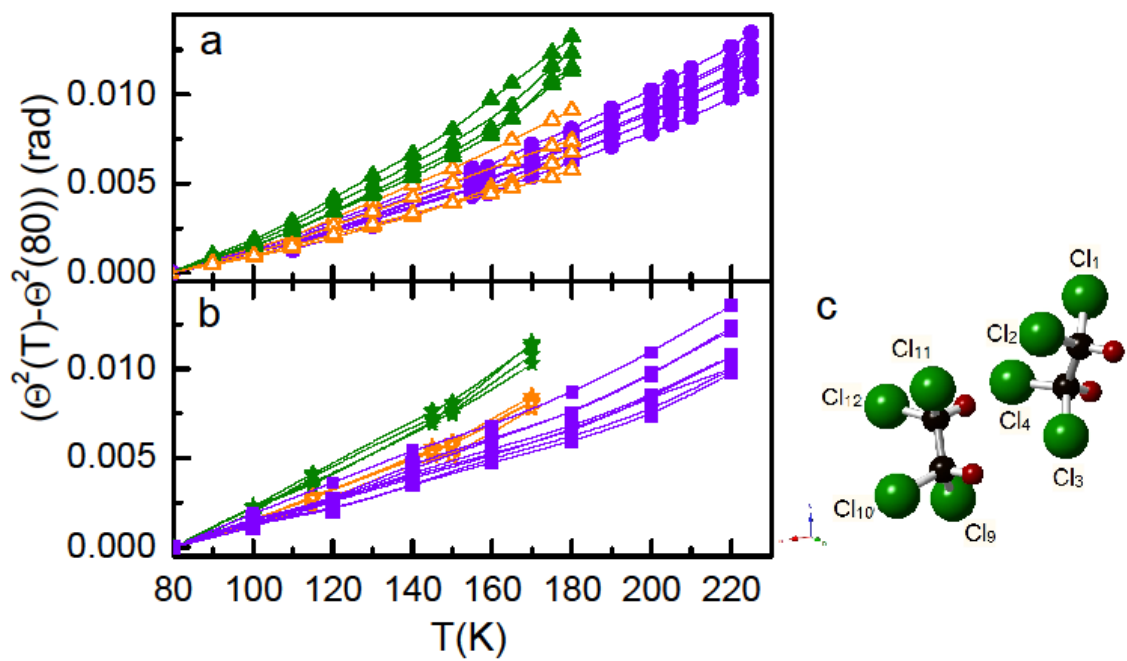

FIG. 5: a) experimental values of $\Delta\theta^2$: $\beta$ phase (●) and $\gamma$ phase (▲,△). b) MDS values of $\Delta\theta^2$: $\beta$ phase (■) and $\gamma$ phase (★, ☆). (★) correspond to $Cl_1$, $Cl_3$ and $Cl_9$, $Cl_{11}$, (☆) correspond to $Cl_2$, $Cl_4$ and $Cl_{10}$, $Cl_{12}$ as shown in c). Atomic labels according to Ref. 5.

In order to unravel which of these situations is responsible for the observed behavior, molecular dynamics simulations were carried out. $\langle \theta^2 \rangle$ was calculated for each of the C-Cl vectors, and the results are shown in Figure 5b. An excellent agreement with experimental NQR data is observed, allowing to associate, beyond question, the two observed $\Delta\theta^2$ groups with the second possible explanation, i.e. in each molecule of the asymmetric unit cell there are two Cl atoms ($Cl_1$, $Cl_3$ and $Cl_9$, $Cl_{11}$) with large $\langle \theta^2 \rangle$ and two ($Cl_2$, $Cl_4$ and $Cl_{10}$, $Cl_{12}$) with small angular displacement. These values are similar for both molecules in the irreducible cell, indicating that both non-equivalent molecules in phase γ have the same vibrational dynamics.

## B. Spin- Lattice relaxation times

The general temperature dependence for the NQR relaxation rate in molecular solids due to torsional motion can be expressed as:[42, 43]

$$\frac{1}{T_1} = AT^\lambda \qquad \lambda \approx 2 \tag{3}$$

Woessner and Gutowsky[42] have also proposed a more general relation for the relaxation rate in solids with symmetric groups, combining torsional motion as well as reorientation (about the preferred axes allowed by symmetry). In such a case, the effective relaxation rate has contributions from both mechanisms, i.e. molecular librations and reorientations:

$$\frac{1}{T_1} = \left(\frac{1}{T_1}\right)_{libr} + \left(\frac{1}{T_1}\right)_{reor}$$

$$\frac{1}{T_1} = AT^\lambda + \frac{\kappa}{\tau_o} e^{-E_a/T} \tag{4}$$

where $A$ and $\kappa/\tau_o$ represent their respective weights; the exponential term represents the contribution due to the reorientation of the molecule or a group in the molecule and $E_a$ is the activation energy for the process. $\kappa$ depends on the kind of reorientation taking place.[44, 45]

For activation energies not too large, the reorientational contribution to the relaxation rate becomes dominant, as it is shown in Figure 6 and, eventually, the resonance line broadens, beyond detection, according to the following expression.[26]

$$\Delta \nu_{reor} = \frac{1}{T_2} = \frac{\overline{\kappa}}{\tau_o} e^{-E_a/T} \quad (5)$$

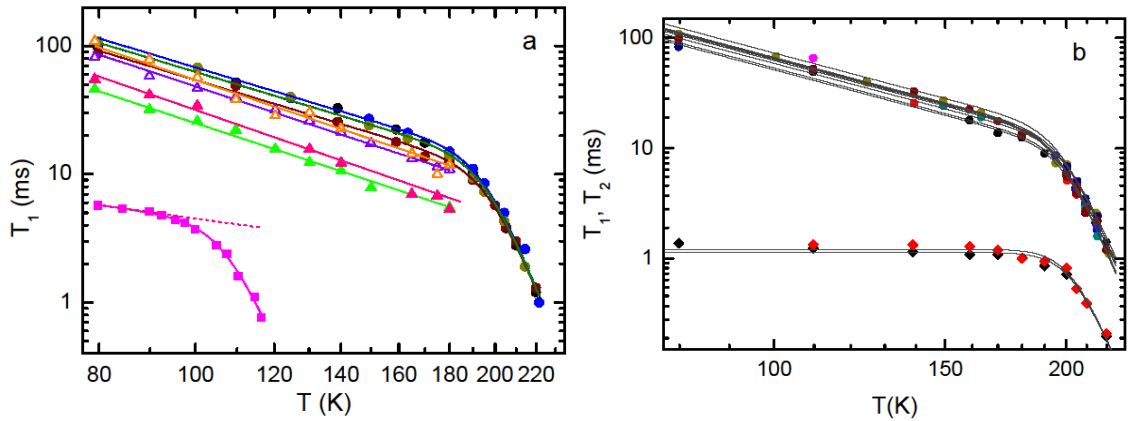

FIG. 6 :(a) Temperature dependence of $T_1$: $\beta$ phase (filled circles), $\gamma$ phase (filled and empty triangles) and glass phase (filled squares). (b) $T_1$ data (filled circles) of the eight lines in $\beta$ phase and $T_2$ data (diamonds) of two representative lines in $\beta$ phase. Continuous lines correspond to the fit of the experimental data with Eqs. (4) and (5). Dashed line shows the $T^{-1}$ behavior of $T_1$ in the glassy phase.

Figure 6a shows the temperature dependence of $T_1$ in the glass, $\gamma$ and $\beta$ phases of deuterated TCE. The temperature dependence of $T_1$ for the eight lines in phase $\beta$ and $T_2$ of two representative lines in this phase are shown in Figure 6b. The first main observation is that $T_1$ in the glass phase is of the order of 16 times lower than $T_1$ in crystalline phase $\beta$ and varies as $T^{-\lambda}$ with $\lambda \sim 1$ below 100 K. This is in good agreement with the strong quadrupolar relaxation observed in amorphous materials as $B_2O_3$, $(Na_2O)_{0.3}(SiO_2)_{0.7}$, $Na_2B_4O_7$[7, 46-48] and in the glassy state of molecular organic systems.[49] The fast nuclear relaxation observed in the glassy phase is attributed to the large amplitude of the EFG fluctuations, caused by localized low-frequency modes. These extra modes, which do not exist in crystals, originate from the disordered structure of glasses and were first invoked to explain their anomalous thermal and acoustic properties.[50] In the glassy state, above 100 K, the presence of a second effective relaxation mechanism is also observed as in glassy orthoterphenyl.[49, 51] Instead, the only relaxation mechanism present in $\gamma$ phase, up to the temperature at which it transforms irreversibly to phase $\beta$, is due to torsional oscillations. It is worth noting that the four lines with stronger temperature dependence are the ones with lower spin-lattice relaxation times. This is due to the fact that molecular oscillations presenting stronger anharmonicity are more effective as relaxation mechanism.

On the other hand, in phase $\beta$ the relaxation mechanism below 200 K is due to lattice vibrations with $T_1$ values comparable to those of phase $\gamma$ corresponding to the lines with the same temperature dependence. Above 200 K, an Arrhenius type relaxation mechanism appears due, probably, to molecular reorientations. It is clear that the reorientational mechanism is the same for the two non-equivalent molecules.

As expected from Eqs. (2), (4), and (5) and experimental evidence in Figure 6b, $T_1$ and $T_2$ values for TCE are indistinguishable when molecular reorientations become dominant.

A least squares fit of data using Eq. (4) gives the parameters shown in Table II.

**Table II:** Least squares fit parameters using Eq. (4).

| Phase | $\lambda$ | $E_a$ (kJmol$^{-1}$) | $\tau_o$ (s) |
|---|---|---|---|
| $\beta$ | 2.3+/-0.1 | 41+/-2 | 4 10$^{-13}$ |
| $\gamma$ | 2.6+/-0.1 | ---- | ---- |
| Glass | 1.1+/-0.3 | 19+/-2 | 4 10$^{-12}$ |

In phase $\beta$ three possible movements could explain the thermally activated behavior of $T_1$: a) internal rotation of the molecule, which must be discarded on the basis of X-ray and MDS results as discussed at the end of section III, b) reorientational motion of the molecule between equivalent potential wells (180º reorientations around the molecular symmetry axis) and c) reorientational motion of the molecule between non-equivalent potential wells.

MDS were carried out in order to determine the kind of reorientations present in phase $\beta$. Two type of reorientations are observed: a position exchange of C, H and Cl atoms in the molecule, associated with a 180º reorientation around the molecular symmetry axis (Figure 7a) and another reorientation between non-equivalent positions (Figure 7b).

Figure 7a clearly shows the exchange between the two carbons and between $Cl_1$ and $Cl_3$ in a particular molecule.

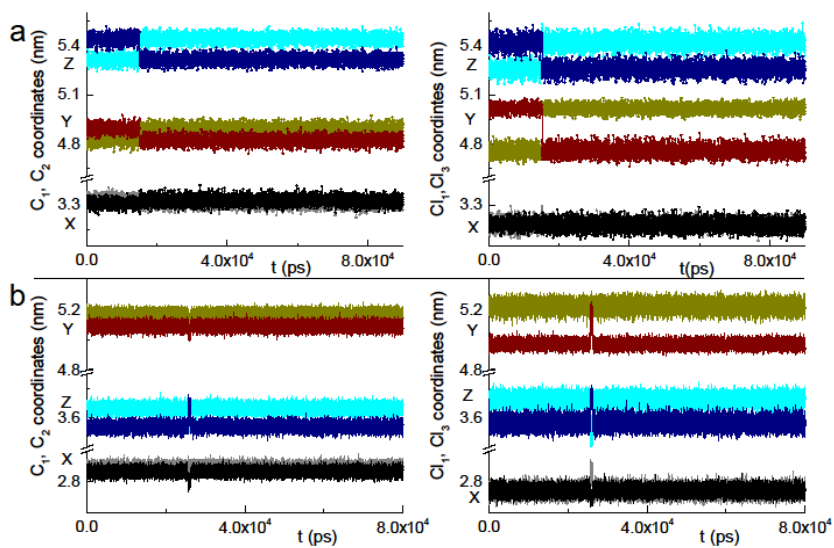

FIG. 7: Carbon and Chlorine coordinates of two different molecules at 230 K, after 300ns of NVT MDS. a) 180º reorientations around the molecular symmetry axis, b) reorientations between non-equivalent positions.

On the other hand, as observed in Figure 7b, in the reorientation between non-equivalent positions, the molecule remains in the new position only a few ns at T $\approx$ 230K.

In this movement the mean angle of reorientation of the different C-Cl bonds, θ (C – Cl$_i$), oscillates between 70º and 90º, as shown in Figure 8.

According to Alexander and Tzalmona´s model for molecular reorientation between equivalent potential wells[44] the factor κ in Eq. (4) is approximately 1. For reorientations between double non-equivalent potential wells, according to Gordeev et al.,[52] κ varies between 0.75 and 0.9 for the angles observed in simulations. Therefore, it is also possible to conclude that $T_1 \sim \tau$. Unfortunately, from experimental data, it is not possible to know which of the reorientations or whether both are responsible for relaxation.

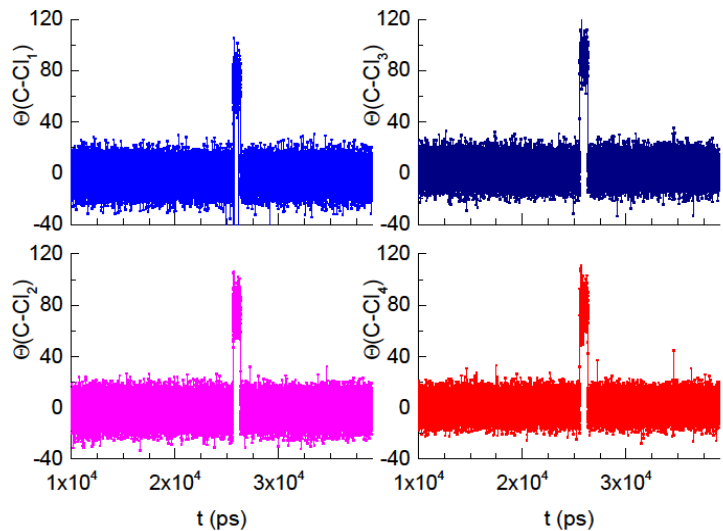

FIG. 8: Reorientation angle of C- Cl$_{(i)}$ bond (i=1- 4).

Figure 9 shows the relaxation map for the glass and phase β of TCE obtained from experimental data together with those reported in literature for Freon 112 (CFCl$_2$-CFCl$_2$).[17] It has been assumed that, in the glass, $T_1 \sim \tau$.

As far as the glass phase is concerned, faster relaxations than the commonly main-relaxation (or α- relaxation) usually appear, among them the so-called Johari-Goldstein (JG) secondary- (or β-) relaxation. This JG relaxation is associated with local, non-cooperative motions of the molecular entities as a whole, thus being different from secondary relaxations due to intramolecular degrees of freedom and found even in single rigid molecules.[21, 53-56] If the Arrhenius type relaxation observed in the glassy state of TCE is considered to be a JG $\beta$-relaxation, then, from the coupling model (CM) relation at $T_g$:[53, 54]

$$\log_{10}(\tau_\beta(T_g)) = \log_{10}(\tau_\alpha(T_g)) - \beta_{KWW}\left(\log_{10}(\tau_\alpha(T_g)) - \log_{10}(t_c)\right) \qquad (6)$$

it would then be possible to obtain the exponent $\beta_{KWW}$ that characterizes the $\alpha$-relaxation in its correlation function given by Kohlrausch-Williams-Watts. Using $t_c \sim 2\ ps$ (crossover time in the CM for small molecular glass formers), $\tau_\alpha(T_g) \approx 100\ s$ and $\log_{10}(\tau_\beta(T_g)) \approx -5$ (obtained from the extrapolation in Figure 9), it results $\beta_{KWW} \cong 0.5$. This result is in good agreement with the experimental values obtained in many glass formers by means of dielectric relaxation spectroscopy.[54]

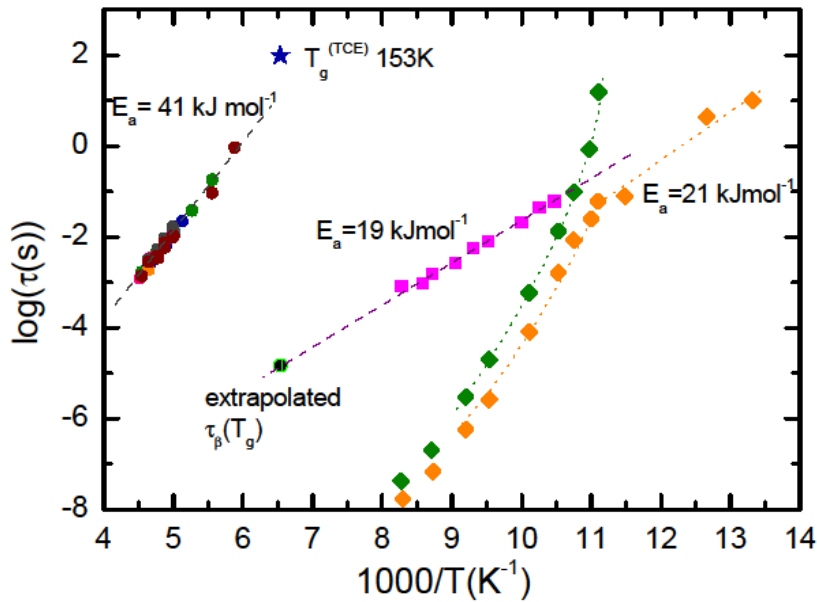

FIG. 9: Relaxation times as a function of temperature. TCE: $\beta$ phase (filled circles), glass phase (■). Dielectric relaxation data for Freon 112 of Pardo *et al.* (Ref. 17) are also shown: (♦) α– relaxation, (♦) β– relaxation.

It is also observed that the activation energies of $\beta$-relaxation in the glass for TCE (19 kJ·mol$^{-1}$) and for Freon 112 (1,1,2,2 tetrachloro-difluoroethane) (21 kJ·mol$^{-1}$) are similar. Moreover, the activation energy for $\beta$-phase (41 kJ/mol) is close to the activation energy for the relaxation in Freon 112 (40 kJ/mol) far from $T_g$ (90 K). In fact, molecular symmetry for both deuterated TCE and Freon 112 are one and the same, the difference being the substitution of Fluorine atoms by lighter Deuterium atoms.

Finally, we would like to emphasize that the glass transition temperature for the liquid state of TCE is 153 K (obtained on heating from DSC measurements at 2 K/min),[5] while the virtual glass transition temperature for the $\beta$ phase, defined as the temperature for which the reorientational movements would reach a characteristic time of 100 s, is around 145 K, so they are actually very close. This would mean that reorientational movements, whatever the phase in which they are developed, are dominant for the glassy dynamics. Such an idea has been brought to light, up to now, only from two experimental evidences, ethanol[57] and a mixed crystal of succinonitrile-glutaronitrile.[58] In both cases, the liquid and the orientationally disordered (plastic) phase can be supercooled and the dynamics can be compared. The results show the outstanding role of the orientational degrees of freedom (those present in the plastic phase) over the translational degrees accompanying the previous ones in the liquid state. In the present case, the on-site orientational motion of TCE molecules offers a new indirect evidence of the prominent role of such orientational disorder.

## V. CONCLUSIONS

In this work, TCE has been dynamically characterized in its crystalline (stable $\beta$ and metastable $\gamma$) phases at normal pressure and in the glass phase through NQR experiments and molecular dynamics simulations.

In the metastable $\gamma$ phase, obtained by crystallization of the supercooled liquid, only molecular librations take place in the temperature range 80-190 K, each molecule containing two Cl atoms with large angular displacements and two with small angular displacements. Instead, in the stable phase $\beta$, molecular reorientations become dominant above 200 K with an activation energy of 41 kJ/mol. From molecular dynamics simulations, two molecular reorientations are possible: a position exchange of C, H and Cl atoms in the molecule, associated with a 180º reorientation around the molecular symmetry axis and a reorientation of the molecule between non-equivalent positions. Unfortunately, from experimental data, it is not possible to know which of the reorientations or if both are responsible for the relaxation observed.

In the glass phase $T_1$ is of the order of 16 times lower that $T_1$ in crystalline phase $\beta$ and varies as $T^{-1}$ below 100 K. This fast nuclear relaxation is attributed to the large amplitude of the EFG fluctuations caused by localized low-frequency modes. Above 100 K, a second effective relaxation mechanism is present, compatible with a molecular reorientational process with an activation energy of 19 kJ/mol. This energy is comparable to that observed in the glass state of a "molecular cousin" compound, Freon 112, for the secondary $\beta$-relaxation. Moreover, the stretching parameter obtained, if the Arrhenius type relaxation observed is assumed to be a JG $\beta$-relaxation, is $\beta_{KWW} \cong 0.5$. This value is in good agreement with the experimental values obtained in many glass formers by means of dielectric relaxation spectroscopy close to the glass transition temperature.

Finally, we would like to stress that, in view of our results, orientational degrees of freedom appearing in "ordered" phases are intrinsically coupled to those present in the glass state of the involved material. Consequently, this work highlights that the study of crystalline phases with some kind of precise disorder can shed light on the secondary relaxations commonly found in the glass state, especially for those originating from non-cooperative molecular motions.

**ACKNOWLEDGEMENTS**

This work was partially supported by CONICET and SECYT-UNC of Argentina and by the Spanish Ministry of Science and Innovation (Grant FIS2014-54734-P) and the Catalan Government (Grant 2014 SGR-581). The MDS were done in Mendieta Cluster from CCAD-UNC, which is part of SNCAD-MinCyT, Argentina.